\documentclass{article}

\usepackage[preprint]{neurips_2026}

\usepackage[utf8]{inputenc} 
\usepackage[T1]{fontenc}    
\usepackage{hyperref}       
\usepackage{url}            
\usepackage{booktabs}       
\usepackage{amsfonts}       
\usepackage{nicefrac}       
\usepackage{microtype}      
\usepackage{xcolor}         
\usepackage{graphicx}
\usepackage{enumitem}
\usepackage{amsmath}
\usepackage{wrapfig}
\usepackage[table]{xcolor}
\usepackage{algorithm}
\usepackage{algorithmicx}
\usepackage{algpseudocode}
\usepackage{multirow}
\usepackage{listings}
\usepackage{textcomp}
\usepackage{framed}
\usepackage[most]{tcolorbox}
\usepackage{titletoc}

\definecolor{promptbg}{rgb}{0.97, 0.98, 1.0}
\definecolor{rulebg}{rgb}{0.98, 1.0, 0.98}
\definecolor{promptkeyword}{rgb}{0.0, 0.2, 0.7}
\definecolor{collabHighlight}{HTML}{1f77b4}
\definecolor{userHighlight}{HTML}{ff7f0e}
\definecolor{featurecolor}{rgb}{0.0, 0.3, 0.5}
\definecolor{commentcolor}{rgb}{0.4, 0.4, 0.4}

\lstdefinestyle{promptstyle}{
    backgroundcolor=\color{promptbg},
    basicstyle=\ttfamily\small,
    breaklines=true,
    columns=fullflexible,
    keepspaces=true,
    showstringspaces=false,
    frame=tb,
    framerule=0pt,
    framexleftmargin=1em,
    framexrightmargin=1em,
    captionpos=b,
    moredelim=[is][\color{promptkeyword}]{@@}{@@}
}

\lstdefinestyle{rulesstyle-plain}{
    backgroundcolor=\color{rulebg},
    basicstyle=\ttfamily\small,
    breaklines=true,
    columns=fullflexible,
    keepspaces=true,
    showstringspaces=false,
    frame=none,
    numbers=none,
    xleftmargin=1em,
    mathescape=true,
    literate={_}{{\_}}1
}

\newtcolorbox{promptbox}[1][]{
  colback=gray!5!white,
  colframe=gray!75!black,
  title={\textbf{#1}},
  fonttitle=\sffamily,
  boxrule=0.8pt,
  arc=2pt,
  left=5pt, right=5pt, top=5pt, bottom=5pt,
  enhanced,
  breakable
}

\titlecontents{section}[0em]
{\addvspace{0.4em}\bfseries}
{\contentslabel{1.5em}}
{}
{\titlerule*[0.7pc]{.}\contentspage}

\titlecontents{subsection}[2em]
{\addvspace{0.1em}\small}
{\contentslabel{2.0em}}
{}
{\titlerule*[0.7pc]{.}\contentspage}

\title{\textsc{Sager}: Self-Evolving User Policy Skills for Recommendation Agent}

\author{%
  \textbf{Zhen Tao$^{\spadesuit}$, Riwei Lai$^{\clubsuit}$, Chenyun Yu$^{\diamondsuit}$\thanks{Corresponding author.}, Weixin Chen$^{\clubsuit}$, Li Chen$^{\clubsuit}$,} \\
  \textbf{Beibei Kong$^{\heartsuit}$, Lei Cheng$^{\heartsuit}$, Chengxiang Zhuo$^{\heartsuit}$, Zang Li$^{\heartsuit}$, Qingqiang Sun$^{\spadesuit}$\footnotemark[1]} \\[4pt]
  \normalfont{$^{\spadesuit}$Great Bay University}   
  \normalfont{$^{\clubsuit}$Hong Kong Baptist University} \\
  \normalfont{$^{\diamondsuit}$Sun Yat-Sen University} \\
  \normalfont{$^{\heartsuit}$Platform and Content Group, Tencent} \\[3pt]
  \normalfont{\texttt{zhentao.tz@gmail.com, \{csrwlai, cswxchen, lichen\}@comp.hkbu.edu.hk}} \\
  \normalfont{\texttt{\{echokong, raycheng, felixzhuo, gavinzli\}@tencent.com}} \\
  \normalfont{\texttt{yuchy35@mail.sysu.edu.cn, qqsun@gbu.edu.cn}}
}

\begin{document}

\maketitle

\begin{abstract}
Large language model (LLM) based recommendation agents personalize \textit{what} they know through evolving per-user semantic memory, yet \textit{how} they reason remains a universal, static system prompt shared identically across all users. This asymmetry is a fundamental bottleneck: when a recommendation fails, the agent updates its memory of user preferences but never interrogates the decision logic that produced the failure, leaving its reasoning process structurally unchanged regardless of how many mistakes it accumulates.
To address this bottleneck, we propose \textbf{\textsc{Sager}} (\textbf{S}elf-Evolving \textbf{AG}ent for P\textbf{E}rsonalized \textbf{R}ecommendation), the first recommendation agent framework in which each user is equipped with a dedicated \textit{policy skill}, a structured natural-language document encoding personalized decision principles that evolves continuously through interaction. \textsc{Sager} introduces a two-representation skill architecture that decouples a rich evolution substrate from a minimal inference-time injection, an incremental contrastive chain-of-thought engine that diagnoses reasoning flaws by contrasting accepted against unchosen items while preserving accumulated priors, and skill-augmented listwise reasoning that creates fine-grained decision boundaries where the evolved skill provides genuine discriminative value. Experiments on four public benchmarks demonstrate that \textsc{Sager} achieves state-of-the-art performance, with gains orthogonal to memory accumulation, confirming that personalizing the reasoning process itself is a qualitatively distinct source of recommendation improvement.
\end{abstract}

\section{Introduction}

Large language model (LLM) based recommendation agents represent the most capable instantiation of personalized recommendation to date~\cite{wu2024survey,zhang2025survey}. By treating recommendation as an agentic reasoning process, these systems maintain per-user semantic memories, retrieve relevant context at inference, and leverage the world knowledge of large language models to produce grounded, interpretable outputs~\cite{zhao2024let,xu2025iagent}. The most recent wave of work pushes this further: collaborative memory agents enrich individual memories with cross-user signals from dynamic semantic graphs~\cite{chen2026memrec}, while self-evolving systems deploy LLM agents to autonomously optimize recommendation models and reward functions at the system scale~\cite{wang2026self,kim2026self}. The field is converging on a compelling vision: a recommendation agent that continuously improves through experience. 

\begin{figure}[t]
    \centering
    \includegraphics[width=0.9\linewidth]{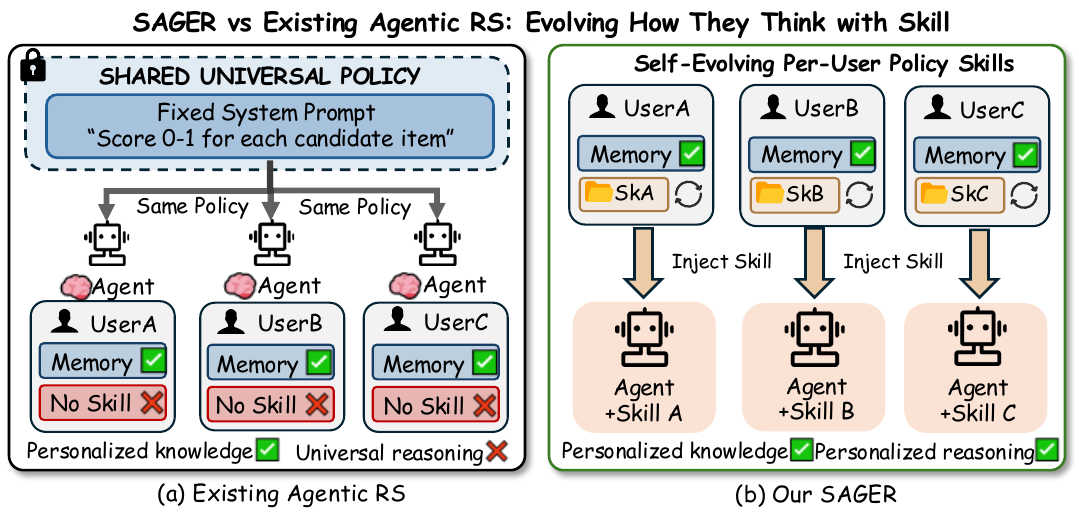}
    \vspace{-1mm}
    \caption{
    \textbf{(a) Existing recommendation agents} personalize \textit{what they know} (memory) but reason about every user with a fixed, shared policy.
    \textbf{(b) \textsc{Sager}} equips each user with a self-evolving \textit{policy skill} that personalizes \textit{how} the agent reasons, not just \textit{what} it remembers.    }
    \label{fig:comparison}
\end{figure}

We share this vision, and observe that a key dimension of improvement remains largely unexplored. Current agents, including the self-evolving ones, focus on enriching what the agent \textit{knows} about each user: building richer memories and retrieving more relevant context. Yet the way the agent \textit{thinks}, specifically the reasoning strategy that translates memory into rankings, typically remains fixed and shared across all users. We hypothesize that this reasoning gap constitutes a significant bottleneck: for users whose decision logic departs from the population mean, richer memories may not be sufficient to close the residual performance gap, because the divergence resides in the \textit{strategy}, not the \textit{facts}.
Concretely, as illustrated in Figure~\ref{fig:comparison}(a), existing recommendation agents operate under a single, shared recommendation \textit{policy}: a fixed set of decision principles hardcoded as a universal system prompt for every user. While these agents produce different recommendations for different users by conditioning on different memories, they reason about every user in the same way, a regime we call \textbf{one-policy-fits-all}. Worse still, when a recommendation fails, the agent updates its memory, recording that an item was not clicked, but it never asks the more consequential question: \textit{what was wrong with the reasoning that produced this recommendation?} The agent knows \textit{what} happened but it never learns \textit{why it was wrong}. Closing this gap calls for a fundamentally new form of adaptation, not memory evolution but \textit{policy evolution}: updating \textit{how} the agent reasons for each individual user.

Realizing per-user, self-evolving recommendation policies, however, is far from straightforward. The seemingly natural approach of maintaining a policy per user and updating it with an LLM after each interaction encounters three fundamental obstacles, each demanding a carefully designed mechanism:

\begin{itemize}[leftmargin=*]
    \item \textbf{The Injection Paradox.} A per-user policy must be rich enough to encode meaningful decision patterns across preferences, behavioral tendencies, and ranking criteria. Yet injecting this rich policy into the LLM's context at inference creates a counter-intuitive tension: enriching the policy representation beyond a narrow threshold actually \textit{degrades} recommendation quality, as the model's attention is diluted away from the immediate instruction and collaborative signals.

    \item \textbf{Evolution Stability under Sparse, Asymmetric Feedback.} The most natural evolution strategy, having the LLM rewrite the policy after each interaction, is fundamentally flawed: a single click provides a noisy, partial signal, and a non-click is deeply ambiguous (the user may have not seen the item, or simply been undecided). Naive full-replacement discards the statistical priors accumulated across many rounds, collapsing a well-calibrated policy into a single-round snapshot. 

    \item \textbf{Cold-Start Initialization and Policy Activation.} A personalized policy cannot exist before any interaction, yet the agent must perform well from the first encounter. Without a principled prior, early evolution is unguided and risks degenerate collapse. Moreover, even a well-formed policy faces a structural \textit{activation barrier}: if the underlying scoring mechanism compresses diverse candidates into tied scores (as conventional pointwise approaches frequently do), the policy finds no decision boundary at which to provide discriminative value, neutralizing its effect precisely when personalization matters most.
\end{itemize}

To address these challenges, we propose \textbf{\textsc{Sager}} (\textbf{S}elf-Evolving \textbf{AG}ent for P\textbf{E}rsonalized \textbf{R}ecommendation), a framework in which each user is equipped with a dedicated \textit{policy skill}, a structured natural-language document encoding personalized decision principles that evolves continuously through interaction feedback. \textsc{Sager} resolves the Injection Paradox through a \textit{two-representation architecture}: a comprehensive full skill repository serves as the evolution substrate, while a principled \textit{slim extraction mechanism} distills it into a minimal-footprint injection that preserves discriminative signals and compresses generic content. Policy evolution is handled by an \textit{incremental contrastive chain-of-thought engine} that outputs structured diffs rather than full replacements, reasoning over the contrast between accepted and unchosen items to isolate specific reasoning flaws while preserving accumulated statistical priors. A \textit{dual-level initialization}, combining population-level policy templates with a zero-cost statistical skill builder, ensures every user begins with a principled prior rather than a blank slate. To overcome the activation barrier, \textsc{Sager} replaces pointwise scoring with \textit{skill-augmented listwise reasoning} that forces strict total orderings over candidates, creating fine-grained decision boundaries where the slim skill can provide genuine discriminative value as a soft tie-breaker.

A key implication of this design is that policy evolution should produce gains \textit{orthogonal} to memory accumulation: improving \textit{how} the agent reasons and improving \textit{what} it knows are independent sources of personalization that compound rather than substitute. Extensive experiments on four diverse benchmarks confirm this hypothesis, with \textsc{Sager} achieving state-of-the-art performance and ablation studies demonstrating that each component contributes independently. 

Our main contributions are as follows:
\begin{itemize}[leftmargin=*]
    \item We identify a structural limitation shared by all existing LLM-based recommendation agents: the recommendation policy is universal and static across users. We formally distinguish \textit{policy evolution} from memory evolution and model evolution, establishing it as a previously unaddressed dimension of personalized recommendation.

    \item We propose \textbf{\textsc{Sager}}, addressing this limitation through three tightly coupled mechanisms: a two-representation policy skill architecture that resolves the Injection Paradox, an incremental contrastive chain-of-thought engine that ensures evolution stability under sparse feedback, and skill-augmented listwise reasoning that overcomes the activation barrier to unlock the policy's discriminative potential.

    \item Comprehensive experiments on four benchmarks demonstrate that \textsc{Sager} consistently outperforms state-of-the-art agents, with ablation studies confirming that policy evolution produces gains orthogonal to memory accumulation.
\end{itemize}

\section{The \textsc{Sager} Framework}
\label{sec:methodology}
\begin{figure*}[t]
    \centering
    \includegraphics[width=1.00\textwidth]{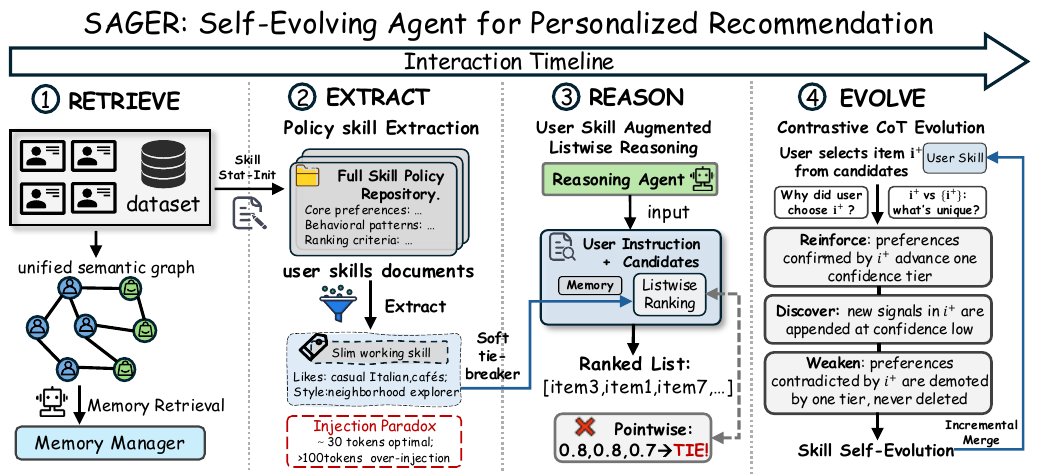}
    \vspace{-3mm}
    \caption{The \textbf{\textsc{Sager}} framework operates in four stages per interaction: \textcircled{1}~\textbf{Retrieve}: the Memory Manager ($\text{LLM}_\text{Mem}$) curates top-$k$ neighbors from the semantic graph and synthesizes collaborative facets $M_{\text{collab}}$; \textcircled{2}~\textbf{Extract}: the full policy skill repository $\mathcal{S}_u^{(t-1)}$ is distilled by $\text{LLM}_{\text{Ext}}$ into a slim working skill $\mathcal{S}_u^{(t-1),\text{slim}}$ ($\sim$30 tokens), respecting the \textit{Cognitive Injection Boundary}; \textcircled{3}~\textbf{Reason}: the Reasoning Agent ($\text{LLM}_\text{Rec}$) performs skill-augmented listwise ranking over $M_{\text{collab}}$, $\mathcal{S}_u^{(t-1),\text{slim}}$ (as soft tie-breaker), and candidates, eliminating the \textit{Score-Tie Bottleneck}; \textcircled{4}~\textbf{Evolve}: a three-phase contrastive CoT engine (Reinforce $\to$ Discover $\to$ Weaken) incrementally refines the policy skill via structured diffs, closing the self-improvement loop.}
    \label{fig:sager_model}
\end{figure*}

\textsc{Sager} equips each user with a dedicated, self-evolving \textit{policy skill}: a structured natural-language document that personalizes not only \textit{what} the agent knows about a user, but \textit{how} it reasons on that user's behalf.
As illustrated in Figure~\ref{fig:sager_model}, each interaction proceeds in four stages: \textbf{Retrieve} collaborative facets from the semantic graph (\S\ref{sec:collab_backbone}), \textbf{Extract} the policy skill into a minimal-footprint
representation (\S\ref{sec:skill_injection}), \textbf{Reason} via skill-augmented listwise ranking (\S\ref{sec:listwise}), and \textbf{Evolve} the policy skill from contrastive feedback (\S\ref{sec:cot_evolution}).

\subsection{Problem Formulation}
\label{sec:problem}

Let $\mathcal{U}$ and $\mathcal{I}$ denote the sets of users and items. Each user $u \in \mathcal{U}$ has an interaction sequence $H_u$ and is associated with a \textit{semantic memory} $M_u$ (an evolving textual narrative of facets) and a \textit{policy skill} $\mathcal{S}_u$ (a structured document encoding personalized decision principles). Given instruction $I$ and candidate set $C \subseteq \mathcal{I}$, the objective is to produce a ranked permutation $\pi^*$ grounded in both community knowledge and individual decision strategy. 

We distinguish three orthogonal dimensions of evolution in agentic recommendation:
\begin{itemize}
    \item \textbf{Memory evolution}: updating \textit{what the agent knows} ($M_u^{(t-1)} \to M_u^{(t)}$).
    \item \textbf{Model evolution}: updating the shared recommendation model or its parameters.
    \item \textbf{Policy evolution}: updating \textit{how the agent reasons} ($\mathcal{S}_u^{(t-1)} \to \mathcal{S}_u^{(t)}$).
\end{itemize}
Most prior work addresses the first dimension, some the second, but none the third. \textsc{Sager} is the first framework to realize per-user policy evolution.

\subsection{Retrieve: Memory-Augmented Context Synthesis}
\label{sec:collab_backbone}

As established in \S\ref{sec:problem}, memory evolution and policy evolution are orthogonal dimensions of improvement. \textsc{Sager} keeps these dimensions cleanly separated: the policy skill governs \textit{how} the agent reasons, while a dedicated memory backbone supplies \textit{what} it knows. Without loss of generality, we instantiate the memory dimension with the state-of-the-art collaborative memory method~\cite{chen2026memrec}, which has demonstrated strong gains on the same benchmarks we evaluate on.

Concretely, a Memory Manager ($\text{LLM}_{\text{Mem}}$) maintains a unified semantic graph $G = (\mathcal{V}, E)$, $\mathcal{V} = \mathcal{U} \cup \mathcal{I}$, with recency- and frequency-weighted edges. At each inference step, domain-adaptive rules select the top-$k$ neighbors $N'_k$, and $\text{LLM}_{\text{Mem}}$ synthesizes their memories into structured facets $\{F\}$ via prompt $P_{\text{synth}}$:
\begin{equation}
\label{eq:collab_synth}
M_{\text{collab}} = \{F\} \leftarrow
\text{LLM}_{\text{Mem}}\left(\text{Rep}(N'_k) \,\|\,
M_u^{(t-1)} \,\|\, P_{\text{synth}}\right),
\end{equation}
where $\text{Rep}(N'_k)$ denotes a tiered representation of the curated neighborhood: the target user $u$ is grounded by their full accumulated memory $M_u^{(t-1)}$, while each neighbor node in $N'_k$ is represented compactly via condensed signals derived from its memory or recent interactions. The resulting $M_{\text{collab}}$ serves as the primary collaborative context in Phase~3, feeding directly into the Reasoning Agent alongside the policy skill to ground the final ranking.

\subsection{Extract: Skill Initialization and Injection}
\label{sec:skill_injection}

The central construct in \textsc{Sager} is the \textbf{personalized policy skill} $\mathcal{S}_u$: a persistent, structured natural-language document encoding each user's individualized decision principles, including \texttt{Core Preferences}, \texttt{Behavioral Patterns} (how the user balances exploration and exploitation), and \texttt{Ranking Criteria} (what principles govern preference decisions). While memory $M_u$ records \textit{what happened} (e.g., ``User visited Italian restaurants and coffee shops''), the policy skill captures \textit{who the user is as a decision-maker} (e.g., ``\textit{likes}: casual dining, local caf\'{e}s; \textit{style}: practical neighborhood explorer''). 

\paragraph{Dual-Level Initialization.} A principled initialization of policy skills is essential: skills must exist before any personalized interaction has occurred, yet must diverge meaningfully from generic defaults as evidence accumulates. We address this through a \textbf{dual-level} design: (i) for \textit{cold-start users} ($|H_u| = 0$), a Meta-LLM synthesizes domain-level policy templates $\mathcal{S}_G$ encoding population-level decision heuristics; (ii) For \textit{warm users} with interaction history ($|H_u| > 0$), a Statistical Skill Builder (\textsc{StatInit}) mines item metadata via domain-adaptive parsers to construct a lightweight initial skill at zero LLM cost:
\begin{equation}
\label{eq:init}
\mathcal{S}_u^{(0)} = \begin{cases}
\mathcal{S}_G & \text{if } |H_u| = 0 \quad \text{(cold-start)}, \\
\text{StatInit}(H_u) & \text{if } |H_u| > 0 \quad \text{(warm)}.
\end{cases}
\end{equation}
From this starting point, the skill self-evolves through interaction: $\mathcal{S}_u^{(t-2)} \to \mathcal{S}_u^{(t-1)}$ via the contrastive CoT engine (\S\ref{sec:cot_evolution}), progressively diverging from the initial prior toward a fully personalized decision profile.

\paragraph{Cognitive Injection Boundary.} Once initialized or evolved, the full policy skill must be injected at each inference step. This operation harbors a counter-intuitive tension we term the \textbf{Injection Paradox}: injecting \textit{more} policy knowledge \textit{degrades} recommendation quality beyond a sharp threshold. Through systematic ablations on \texttt{gpt-4o-mini} (\S\ref{sec:skill_analysis}), we find that injection quality is non-monotonic in length: a $\sim$30 token distillation achieves the Pareto-optimal point, while beyond $\sim$100 tokens, \textit{attention dilution} sets in as population-generic terms crowd out the immediate instruction and collaborative facets. We term this threshold the \textit{Cognitive Injection Boundary}. Note that the exact boundary may shift for models with different context-handling characteristics. \footnote{This boundary is model-specific; LLMs with stronger long-context retrieval capabilities may tolerate longer skill injections; We report the empirical boundary for \texttt{gpt-4o-mini}, the underlying non-monotonic trend, however, is expected to generalize.}

This finding motivates a \textbf{two-representation} design. The \textit{Full Skill Repository} ($\mathcal{S}_u^{(t-1)}$, $\sim$1500 chars) is maintained on disk as the substrate for evolution. At each inference step, a \textit{Slim Working Skill} ($\mathcal{S}_u^{(t-1),\text{slim}}$, $\sim$30 tokens) is extracted via prompt $P_{\text{extract}}$:
\begin{equation}
\label{eq:slim_extract}
\mathcal{S}_u^{(t-1),\text{slim}} \leftarrow \text{LLM}_{\text{Ext}}
(\mathcal{S}_u^{(t-1)}\,\| \, P_{\text{extract}}).
\end{equation}
$\text{LLM}_{\text{Ext}}$ produces \texttt{likes} (2--3 dominant preference themes) and \texttt{style} (one decision-pattern phrase), e.g., \textit{``likes: casual Thai, spicy food | style: budget-friendly''}. The resulting $\mathcal{S}_u^{(t-1),\text{slim}}$ is then injected into Phase~3, providing high-signal personalized guidance without triggering the Injection Paradox.

\subsection{Reason: Skill-Augmented Listwise Ranking}
\label{sec:listwise}

Phase~3 integrates the collaborative facets $M_{\text{collab}}$ from Phase~1 and the slim policy skill $\mathcal{S}_u^{(t-1),\text{slim}}$ from Phase~2 to produce the final ranking.

\paragraph{Score-Tie Bottleneck.} In conventional pointwise scoring, the LLM assigns an independent score to each candidate. Since LLMs inherently struggle with absolute calibration, diverse candidates are compressed into a narrow score range, producing tied top scores in a substantial fraction of predictions. We term this the \textbf{Score-Tie Bottleneck}. Therefore, we replace pointwise scoring with listwise ranking, which forces a strict total ordering and eliminates ties by construction. This choice is further grounded in two complementary properties: under the Plackett-Luce model~\cite{luce1959individual}, a ranking is generated by sequentially selecting the most preferred item from the remaining candidates, explicitly capturing inter-item dependencies that pointwise scoring ignores; and listwise judgment aligns with LLMs' well-documented comparative reasoning advantage, asking the model to judge \textit{which} item is better rather than \textit{how good} each item is in isolation~\cite{hou2024large}.

\paragraph{Skill-Augmented Ranking.} With ties eliminated, $\mathcal{S}_u^{(t-1),\text{slim}}$ can now serve as an effective soft tie-breaker. When candidates are close in quality, the model must actively consult $\mathcal{S}_u^{(t-1),\text{slim}}$ to commit to a strict ordering, activating it as a genuine decision signal rather than passive context. Concretely, the Reasoning Agent produces $\pi^*$ as:
\begin{equation}
\label{eq:listwise}
\pi^* = \text{LLM}_{\text{Rec}}\left(
I \;\|\; M_{\text{collab}} \;\|\;
\mathcal{S}_u^{(t-1),\text{slim}} \;\|\; C \;\|\; P_{\text{list}}\right),
\end{equation}
where $P_{\text{list}}$ encodes a three-tier priority: instruction match $\succ$ facets alignment $\succ$ slim skill tie-breaking.

\subsection{Evolve: Incremental Contrastive CoT}
\label{sec:cot_evolution}

After Phase~3 produces ranking $\pi^*$ and the user selects item $i^+$, Phase~4 uses this interaction signal to evolve the policy skill. A static policy, however well-initialized, cannot capture shifting preferences. \textsc{Sager} refines $\mathcal{S}_u^{(t-1)}$ via an \textbf{Incremental Contrastive CoT} engine built around a \textit{refine-not-replace} principle: a single interaction should shift \textit{confidence estimates} of existing preferences, not reconstruct the entire policy.

\paragraph{Contrastive Input Construction.} Given user $u$'s positive selection $i^+$ from $N$ candidates $C = \{i^+\} \cup \{i^-_1, \ldots, i^-_{N-1}\}$, the remaining $N{-}1$ items form the unchosen set $\{i^-\}$. Inspired by the curriculum learning principle~\cite{bengio2009curriculum}, the CoT engine organizes the unchosen set in \textit{reverse ranking order} (lowest-ranked by $\pi^*$ first) as contrastive input. This progression moves from easy contrasts (items clearly dissimilar to $i^+$) to diagnostically rich near-misses (items the model nearly ranked first), concentrating the strongest learning signal at the end of the analysis.

\paragraph{Three-Phase Contrastive Reasoning.} Against this contrastive input and the existing skill $\mathcal{S}_u^{(t-1)}$, the engine executes three focused phases:

\noindent\textsc{Reinforce.} Which existing preferences (patterns, or criteria) does $i^+$ confirm? A preference is \textit{reinforced} if $i^+$ clearly exemplifies it (e.g., the skill encodes ``casual Italian dining'' and $i^+$ is an Italian trattoria). Reinforcement advances the attribute by one confidence tier (low$\,\to\,$medium$\,\to\,$high). Only preferences at \texttt{medium} or above are eligible for the slim skill, making this the critical step that translates discovered signals into actionable guidance.

\noindent\textsc{Discover.} What new preferences (patterns, or criteria) emerge from $i^+$ that are absent from $\mathcal{S}_u^{(t-1)}$? New preferences are appended at confidence \texttt{low} as single-observation hypotheses, preventing overfitting while recording exploratory signals. Crucially, unchosen $\neq$ disliked: no avoidance rules are extracted from non-clicks, consistent with the \textit{Positive Exploration Paradigm}: skills encode exclusively positive signals (\texttt{likes} and \texttt{style}).

\noindent\textsc{Weaken.} Only if $i^+$ clearly \textit{contradicts} an existing preference (pattern, or criterion) is that preference demoted by exactly one tier (high$\,\to\,$medium, or medium$\,\to\,$low), never deleted outright. This conservative demotion ensures that statistically robust priors from \textsc{StatInit} require multiple contradictions before being discarded.

\paragraph{Incremental Merge Protocol.} The contrastive reasoning output is a structured diff $\Delta_u = \{\texttt{reinforced},\;\texttt{discovered},\;\texttt{weakened}\}$, merged into $\mathcal{S}_u^{(t-1)}$ via confidence-tier updates. High-confidence preferences from \textsc{StatInit} are \textit{protected} and require multiple contradictions before being downgraded. The resulting update is:
\begin{equation}
\label{eq:cot_update}
\mathcal{S}_u^{(t)} \leftarrow \text{Merge}\!\left(
\mathcal{S}_u^{(t-1)},\;
\text{CoT}_{\text{incr}}\!\left(i^+,\;\{i^-\},\;
M_{\text{collab}},\;
\mathcal{S}_u^{(t-1)}\right)\right).
\end{equation}

Full prompt templates are in Appendix~\ref{sec:app_prompts}. The complete per-interaction procedure is summarized in Algorithm~\ref{alg:sager_inference}.

\begin{algorithm}[t]
\caption{\textsc{Sager}: Inference and Evolution Pipeline}
\label{alg:sager_inference}
\small
\begin{algorithmic}[1]
\Require User $u$, neighbors $k$, instruction $I$, candidates $C$,
graph $G$, skill $\mathcal{S}_u^{(t-1)}$, memory $M_u^{(t-1)}$
\Ensure Ranked permutation $\pi^*$, updated $M_u^{(t)}$, updated $\mathcal{S}_u^{(t)}$

\Statex \hfill \textit{\% \textbf{Phase 1: Retrieve}}
\State $N'_k \gets \textsc{DomainPrune}(G, u, k)$
\State $M_{\text{collab}} \gets \textsc{QualityFilter}(
  \text{LLM}_{\text{Mem}}(N'_k \;\|\; M_u^{(t-1)} \;\| \;
  P_{\text{synth}}))$

\Statex \hfill \textit{\% \textbf{Phase 2: Extract}}
\State $\mathcal{S}_u^{(t-1),\text{slim}} \gets \textsc{SlimExtract}(
  \mathcal{S}_u^{(t-1)}, P_{\text{extract}})
  \oplus \text{``tie-breaker only''}$

\Statex \hfill \textit{\% \textbf{Phase 3: Reason}}
\State $\pi^* \gets \text{LLM}_{\text{Rec}}(
  I \;\|\; M_{\text{collab}} \;\|\; \mathcal{S}_u^{(t-1),\text{slim}} \;\|\;
  C \;\|\; P_{\text{list}})$

\Statex \hfill \textit{\% \textbf{Phase 4: Evolve}}
\State $M_u^{(t)}, M_{i_c}^{(t)}, \{\Delta M_j\} \gets
  \textsc{Propagate}(i_c, M_u^{(t-1)}, N'_k)$
\State $\Delta_u \gets \textsc{CoT-Incr}(
  i^+,\, \textsc{ReverseRank}(\pi^*) \setminus \{i^+\},\,
  M_{\text{collab}},\, \mathcal{S}_u^{(t-1)})$
\State $\mathcal{S}_u^{(t)} \gets \textsc{Merge}(
  \mathcal{S}_u^{(t-1)}, \Delta_u)$
\State \Return $\pi^*$, $M_u^{(t)}$, $\mathcal{S}_u^{(t)}$
\end{algorithmic}
\end{algorithm}

\section{Empirical Evaluation}
\label{sec:experiments}

We conduct extensive experiments to answer the following research questions:
\begin{itemize}[leftmargin=*]
    \item \textbf{RQ1 (Overall Performance):} Does \textsc{Sager} outperform state-of-the-art traditional and agentic baselines across diverse benchmarks? Does policy evolution yield gains orthogonal to memory accumulation?

    \item \textbf{RQ2 (Generalizability):} Is \textsc{Sager}'s policy skill mechanism a model-agnostic plug-in that benefits agents with fundamentally different memory paradigms?

    \item \textbf{RQ3 (Component Ablation):} How does each core component (policy skill, \textsc{StatInit}, listwise reasoning, incremental CoT) independently contribute?

    \item \textbf{RQ4 (Design Sensitivity):} How sensitive is \textsc{Sager} to key design choices, including warmup rounds $\tau$ and slim skill injection length?
\end{itemize}

\subsection{Experimental Setup}

\paragraph{Datasets.}
We evaluate on four widely used benchmark datasets covering diverse domains with varying interaction densities: \textbf{Amazon Books}, \textbf{Amazon Goodreads}, \textbf{MovieTV}, and \textbf{Yelp}. For all datasets, we use the specific user instructions and evaluation splits provided by InstructRec~\cite{xu2025iagent} to ensure fair comparison with instruction-following baselines. Table~\ref{tab:dataset_stats} summarizes the basic statistics; detailed descriptions are provided in Appendix~\ref{app:dataset_details}.

\begin{table}[h!]
\centering
\caption{Statistics of the datasets used in experiments.}
\label{tab:dataset_stats}
\resizebox{0.58\textwidth}{!}{
\begin{tabular}{l|rrrrr}
\toprule
\textbf{Dataset} & $\mathbf{|U|}$ & $\mathbf{|I|}$ & $\mathbf{|E|}$ & $\mathbf{\bar{L}_u}$ & \textbf{Density} \\
\midrule
Books & 7.4K & 120.9K & 207.8K & 28.2 & 2.33e-4 \\
GoodReads & 11.7K & 57.4K & 618.3K & 52.7 & 9.19e-4 \\
MovieTV & 5.6K & 29.0K & 79.7K & 14.1 & 4.87e-4 \\
Yelp & 3.0K & 31.6K & 63.1K & 21.4 & 6.77e-4 \\
\bottomrule
\end{tabular}
}
\end{table}

\begin{table*}[t]
\centering
\caption{Main results for \textbf{Goodreads} and \textbf{Books}. All improvements are statistically significant ($p < 0.05$). \underline{Underline}=best baseline, \textbf{Bold}=best overall.}
\label{tab:main_results_1}
\resizebox{\textwidth}{!}{%
\begin{tabular}{lrrrrrrrrrr}
\toprule
& \multicolumn{5}{c}{\textbf{Goodreads}} & \multicolumn{5}{c}{\textbf{Books}} \\
\cmidrule(lr){2-6} \cmidrule(lr){7-11}
\textbf{Model} & \textbf{H@1} & \textbf{H@3} & \textbf{N@3} & \textbf{H@5} & \textbf{N@5} & \textbf{H@1} & \textbf{H@3} & \textbf{N@3} & \textbf{H@5} & \textbf{N@5} \\
\midrule
LightGCN & 0.2499 & 0.5879 & 0.4432 & 0.7903 & 0.5263 & 0.1753 & 0.3259 & 0.2596 & 0.5703 & 0.3592 \\
SASRec & 0.1324 & 0.3518 & 0.2576 & 0.5407 & 0.3349 & 0.0914 & 0.2830 & 0.2001 & 0.4845 & 0.2824 \\
P5 & 0.1569 & 0.3229 & 0.2509 & 0.5060 & 0.3256 & 0.2192 & 0.3607 & 0.2994 & 0.5273 & 0.3671 \\
\midrule
Vanilla LLM & 0.2864 & 0.4662 & 0.3948 & 0.7390 & 0.5041 & 0.3138 & 0.5617 & 0.4533 & 0.7270 & 0.5226 \\
iAgent & 0.2617 & 0.4949 & 0.3954 & 0.6591 & 0.4626 & 0.3925 & 0.5560 & 0.4858 & 0.6905 & 0.5409 \\
\midrule
RecBot & 0.2705 & 0.4754 & 0.3876 & 0.6495 & 0.4589 & 0.3984 & 0.5491 & 0.4846 & 0.6786 & 0.5376 \\
AgentCF & 0.2951 & 0.5910 & 0.4654 & 0.7726 & 0.5399 & 0.3457 & 0.6060 & 0.4960 & 0.7403 & 0.5512 \\
i$^2$Agent & 0.3099 & 0.6079 & 0.4825 & 0.7675 & 0.5481 & 0.4453 & 0.6517 & 0.5649 & 0.7708 & 0.6138 \\
MemRec & \underline{0.3997} & \underline{0.6658} & \underline{0.5540} & \underline{0.8052} & \underline{0.6112}
       & \underline{0.5117} & \underline{0.6915} & \underline{0.6152} & \underline{0.8007} & \underline{0.6601} \\
\midrule
\rowcolor{gray!20}
\textbf{\textsc{Sager}} & \textbf{0.4593} & \textbf{0.7001} & \textbf{0.5989} & \textbf{0.8273} & \textbf{0.6512}
                       & \textbf{0.5215} & \textbf{0.7046} & \textbf{0.6261} & \textbf{0.8101} & \textbf{0.6694} \\
\bottomrule
\end{tabular}%
}
\end{table*}

\begin{table*}[t]
\centering
\caption{Main results for \textbf{Yelp} and \textbf{MovieTV}. Notation follows Table~\ref{tab:main_results_1}; all improvements are significant ($p < 0.05$).}
\label{tab:main_results_2}
\resizebox{\textwidth}{!}{%
\begin{tabular}{lrrrrrrrrrr}
\toprule
& \multicolumn{5}{c}{\textbf{Yelp}} & \multicolumn{5}{c}{\textbf{MovieTV}} \\
\cmidrule(lr){2-6} \cmidrule(lr){7-11}
\textbf{Model} & \textbf{H@1} & \textbf{H@3} & \textbf{N@3} & \textbf{H@5} & \textbf{N@5} & \textbf{H@1} & \textbf{H@3} & \textbf{N@3} & \textbf{H@5} & \textbf{N@5} \\
\midrule
LightGCN & 0.3444 & 0.5658 & 0.4720 & 0.7546 & 0.5494 & 0.3482 & 0.5643 & 0.4738 & 0.6883 & 0.5241 \\
SASRec & 0.2305 & 0.4312 & 0.3458 & 0.5597 & 0.3980 & 0.3399 & 0.5233 & 0.4470 & 0.6382 & 0.4942 \\
P5 & 0.1444 & 0.3207 & 0.2435 & 0.5220 & 0.4785 & 0.1696 & 0.3206 & 0.2554 & 0.5008 & 0.3290 \\
\midrule
Vanilla LLM & 0.1692 & 0.5275 & 0.3696 & 0.6861 & 0.4360 & 0.4050 & 0.7764 & 0.6098 & 0.8603 & 0.6445 \\
iAgent & 0.3995 & 0.6005 & 0.5148 & 0.7300 & 0.5681 & 0.4253 & 0.6170 & 0.5361 & 0.7420 & 0.5871 \\
\midrule
RecBot & 0.4007 & 0.6003 & 0.5156 & 0.7169 & 0.5636 & 0.4367 & 0.6113 & 0.5375 & 0.7309 & 0.5866 \\
AgentCF & 0.1925 & 0.4374 & 0.3326 & 0.6374 & 0.4147 & 0.3906 & 0.6693 & 0.5523 & 0.7864 & 0.6006 \\
i$^2$Agent & 0.4205 & 0.6454 & 0.5517 & 0.7648 & 0.6007 & 0.4912 & 0.7225 & 0.6262 & 0.8221 & 0.6672 \\
MemRec & \underline{0.4868} & \underline{0.6912} & \underline{0.6053} & \underline{0.7908} & \underline{0.6463}
       & \underline{0.5882} & \underline{0.7819} & \underline{0.7011} & \underline{0.8817} & \underline{0.7422} \\
\midrule
\rowcolor{gray!20}
\textbf{\textsc{Sager}} & \textbf{0.5227} & \textbf{0.7220} & \textbf{0.6388} & \textbf{0.8142} & \textbf{0.6766}
                       & \textbf{0.6284} & \textbf{0.8318} & \textbf{0.7467} & \textbf{0.9092} & \textbf{0.7786} \\
\bottomrule
\end{tabular}%
}
\end{table*}

\paragraph{Baselines.}
We evaluate \textsc{Sager} against a comprehensive suite of baselines grouped by their underlying memory and policy paradigms. The first group comprises traditional pre-LLM methods using dense latent embeddings: \textbf{LightGCN}~\cite{he2020lightgcn}, \textbf{SASRec}~\cite{kang2018self}, and \textbf{P5}~\cite{geng2022recommendation}. The second group encompasses LLM-based approaches with progressively richer memory: (1)~\textit{no explicit memory} (\textbf{Vanilla LLM}~\cite{liu2023chatgpt}); (2)~\textit{static memory} (\textbf{iAgent}~\cite{xu2025iagent}); (3)~\textit{dynamic isolated memory} (\textbf{i$^2$Agent}~\cite{xu2025iagent}, \textbf{AgentCF}~\cite{zhang2024agentcf}, \textbf{RecBot}~\cite{tang2025interactive}); and (4)~\textit{dynamic collaborative memory} (\textbf{MemRec}~\cite{chen2026memrec}), which introduces collaborative graph propagation but operates with a universal, static policy. In contrast, \textsc{Sager} introduces a new paradigm, \textit{per-user self-evolving policy}, where each user's recommendation reasoning is a persistent, evolving skill file that personalizes not only what the agent knows, but how it reasons. Baseline implementation details are provided in Appendix~\ref{app:baseline_details}.

\paragraph{Implementation Details.}
\label{sec:experimental_setup}
We implement \textsc{Sager} using \textbf{gpt-4o-mini}~\cite{openai2024hello} as the LLM. The memory retrieval backbone is built on \textbf{MemRec}~\cite{chen2026memrec}; memory-related hyperparameters follow the MemRec configuration.
For all experiments, we fix the candidate list size at $N=10$ (one positive item and nine randomly sampled negatives). The warmup budget $\tau$ controls how many evolution rounds each user's policy skill undergoes before evaluation; once warmup completes, the skill is \textbf{frozen} so that all test-time predictions reflect a fixed policy, ensuring fair comparison with baselines that do not evolve at inference time. We report \textbf{Hit Rate (H@K)} and \textbf{NDCG (N@K)} for $K \in \{1, 3, 5\}$.
Following~\cite{zhang2024agentcf}, we utilize a randomly sampled subset of 1000 users in the subsequent sensitivity study, with a \textbf{fixed candidate set} generated once per user and reused across all conditions, eliminating variance from random sampling.

Key hyperparameters for \textsc{Sager} policy skills: warmup rounds $\tau \in \{0, 1, 2, 3\}$, with the default set to $\tau{=}2$; slim skill injection length searched over $\{10, 30, 80, 150, 200\}$ tokens, with the default set to 30 tokens near the Cognitive Injection Boundary; confidence tiers have three levels (\texttt{low}/\texttt{medium}/\texttt{high}), with only \texttt{medium} and above eligible for injection. Complete implementation details are in Appendix~\ref{sec:implementation_details}.

\subsection{Main Results (RQ1)}

Tables~\ref{tab:main_results_1} and \ref{tab:main_results_2} present the comprehensive performance comparison across four datasets. All reported improvements are statistically significant ($p < 0.05$). We highlight four key findings.

\begin{itemize}[leftmargin=*]

\item \textbf{\textsc{Sager} achieves consistent state-of-the-art performance.}
\textsc{Sager} outperforms all baselines across the vast majority of metrics on all four benchmarks. The gains are most pronounced on Goodreads (H@1 = 0.4593, \textbf{+14.91\%} over MemRec), followed by Yelp (+7.37\%), MovieTV (+6.83\%), and Books (+1.92\%). Since \textsc{Sager} and MemRec share the identical memory retrieval backbone, these improvements are \textit{directly and exclusively} attributable to the policy skill layer, confirming that per-user policy evolution constitutes an independent personalization dimension orthogonal to memory accumulation.

\item \textbf{The performance hierarchy validates a clear progression of recommendation paradigms.}
Results across agentic baselines broadly follow a progressive hierarchy: no memory (Vanilla LLM) $<$ static memory (iAgent) $<$ dynamic isolated memory (i$^2$Agent, AgentCF) $<$ dynamic collaborative memory (MemRec) $<$ \textbf{self-evolving per-user policy} (\textsc{Sager}). Each layer of sophistication, from memory to collaboration to personalized policy, contributes additive gains. Notably, the \textsc{Sager}$-$MemRec transition is the only one that changes \textit{how} the agent reasons rather than \textit{what} it knows, establishing policy evolution as a qualitatively distinct dimension of improvement.

\item \textbf{Policy evolution primarily sharpens top-1 precision.}
H@1 improvements are systematically larger than H@3/H@5 across all datasets: policy skills provide the greatest discriminative signal at the decision boundary where candidate scores are near-tied, while having limited impact where score gaps are already large. This \textit{asymmetric gain pattern} is consistent with the slim skill's soft tie-breaker design (\S\ref{sec:listwise}): the skill activates most strongly when the model must distinguish between semantically close candidates.

\item \textbf{Gain magnitude correlates with interaction density.}
The gain ordering (Goodreads $+14.9\%$ $>$ Yelp $+7.4\%$ $>$ MovieTV $+6.8\%$ $>$ Books $+1.9\%$) correlates monotonically with dataset density (Table~\ref{tab:dataset_stats}): Goodreads has the highest density ($9.19{\times}10^{-4}$) while Books has the lowest ($2.33{\times}10^{-4}$). We term this the \textit{StatInit density hypothesis}: denser interaction graphs yield higher-quality statistical priors from \textsc{StatInit}, providing a stronger foundation for subsequent CoT evolution. Sparse graphs limit the initial prior quality, constraining the headroom for policy-driven improvement. Additionally, Books lacks structured metadata fields (e.g., explicit genre tags or categorical labels); the domain-adaptive parser must infer genre and mood signals from unstructured free-text descriptions alone, producing lower-coverage and less discriminative priors than domains with rich structured metadata such as Yelp, which further suppresses the ceiling for skill-driven gains.

\end{itemize}

\subsection{Generalizability to Other Agents (RQ2)}
\label{sec:generalizability}

A key claim of \textsc{Sager} is that policy evolution is a modular, agent-agnostic capability. To validate this, we augment three representative baselines covering three distinct paradigms: \textbf{iAgent} (static user profile), \textbf{i$^2$Agent} (dynamic profile update), and \textbf{Vanilla LLM} (memory-free LLM reranking). For each baseline, we inject a per-user evolving skill file initialized via \textsc{StatInit} and updated with Incremental CoT ($\tau{=}1$), while keeping all other components unchanged. This design isolates the contribution of skill evolution from any memory-related factors.

\begin{figure}[t]
    \centering
    \includegraphics[width=\columnwidth]{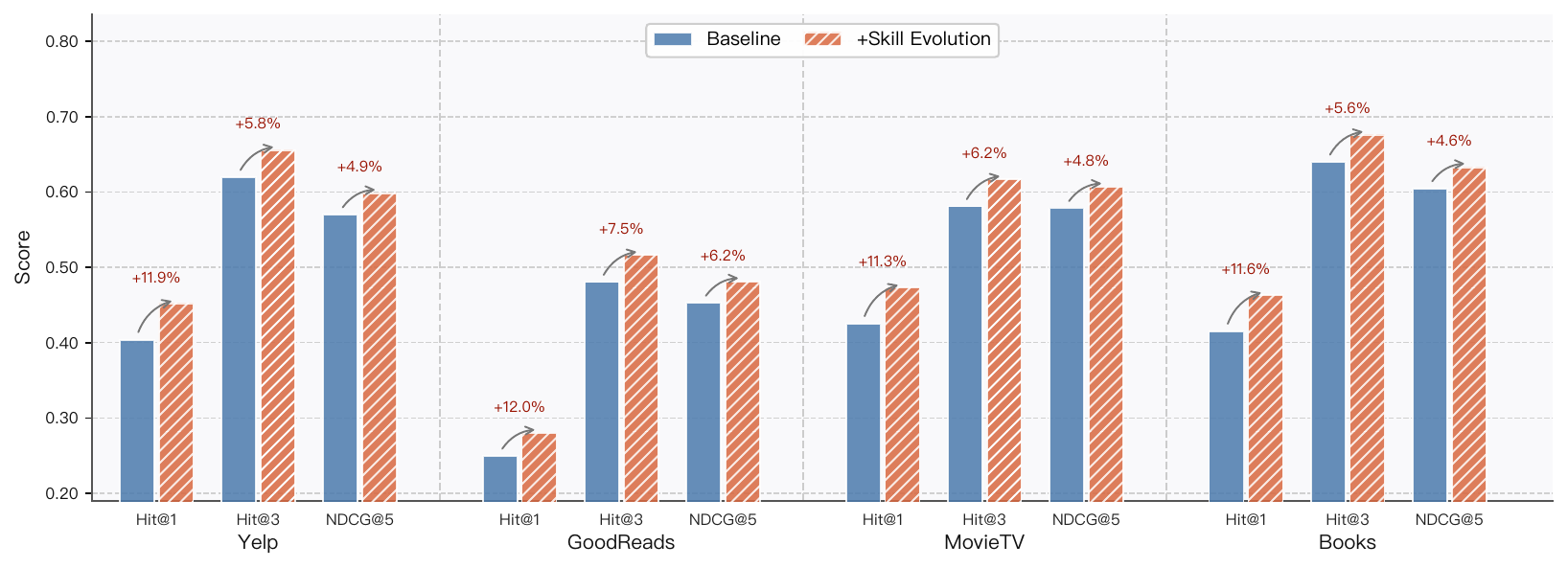}
    \vspace{-3mm}
    \caption{iAgent with and without \textsc{Sager} skill evolution across four datasets. Consistent improvements across all datasets demonstrate that skill evolution is effective even for static-profile agents.}
    \label{fig:gen_iagent_static}
\end{figure}

\begin{figure}[t]
    \centering
    \includegraphics[width=\columnwidth]{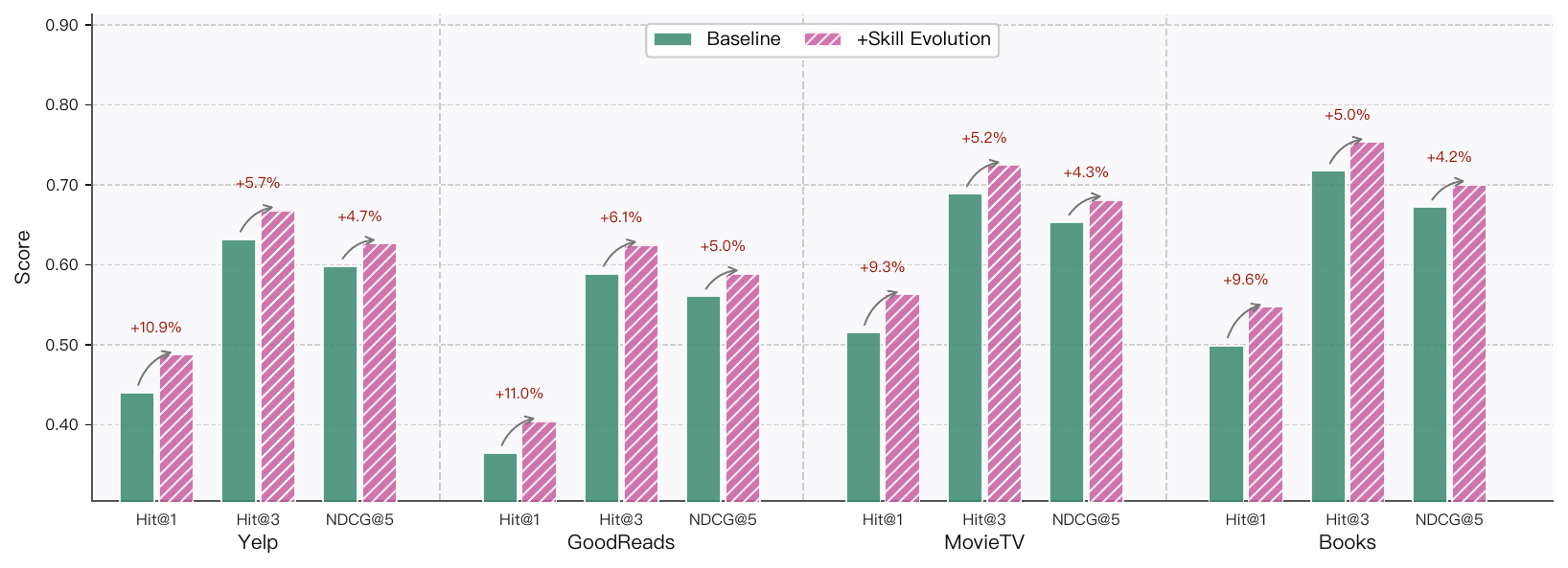}
    \vspace{-3mm}
    \caption{i$^2$Agent with and without \textsc{Sager} skill evolution. Skill evolution provides additive gains on top of dynamic profile updates, confirming orthogonality between profile-level memory and policy-level skill.}
    \label{fig:gen_iagent_dynamic}
\end{figure}

\begin{figure}[t]
    \centering
    \includegraphics[width=\columnwidth]{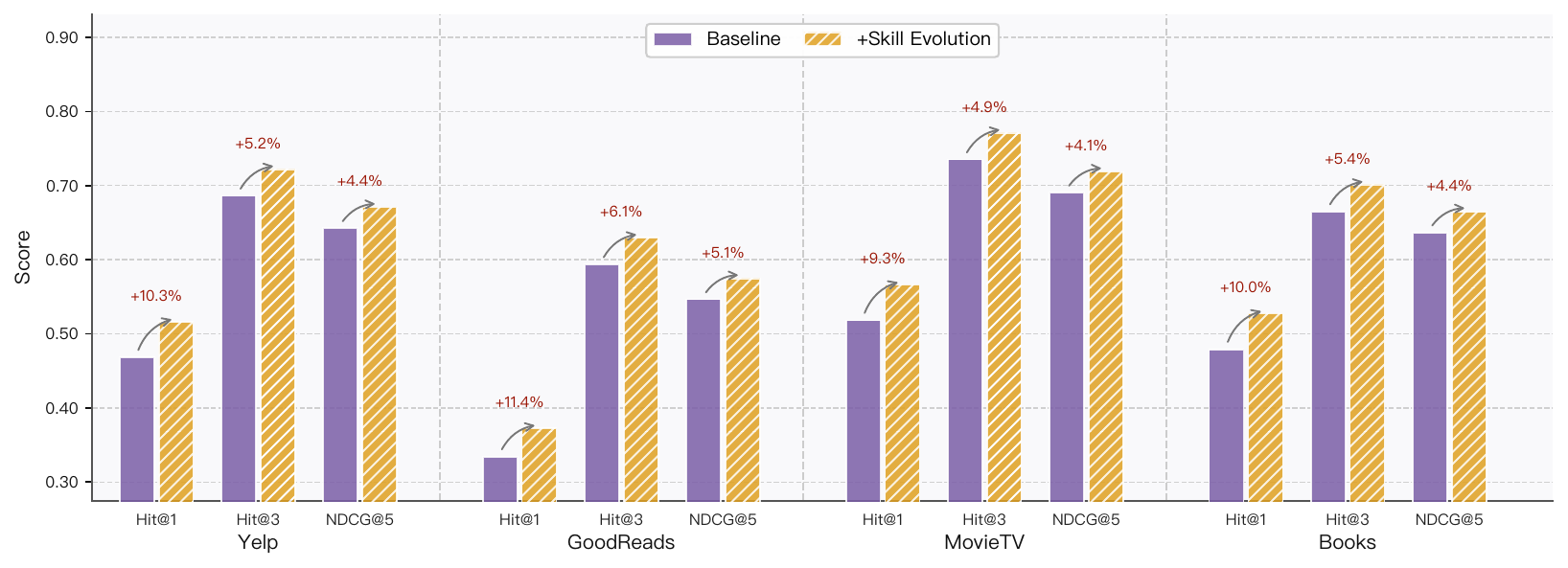}
    \vspace{-3mm}
    \caption{Vanilla LLM with and without \textsc{Sager} skill evolution. Skill evolution yields consistent improvements across all datasets, demonstrating the module's practical utility.}
    \label{fig:gen_llmrec}
\end{figure}

Figures~\ref{fig:gen_iagent_static}--\ref{fig:gen_llmrec} compare each baseline (solid bar) and its skill-augmented variant (hatched bar) on Hit@1, Hit@3, and NDCG@5 across all four datasets. Three observations emerge:

\begin{itemize}[leftmargin=*]
    \item \textbf{iAgent} shows the most pronounced Hit@1 gains, particularly on high-density datasets where \textsc{StatInit} produces richer priors. This confirms that skill evolution is most effective when the statistical initialization is well-grounded, consistent with the density hypothesis observed in main results.
    \item \textbf{i$^2$Agent} benefits from skill evolution \textit{on top of} its dynamic profile updates. Since i$^2$Agent already adapts \textit{what} it knows per interaction, the additional gain from skill injection demonstrates that personalizing \textit{how} the agent reasons is genuinely orthogonal to memory-level personalization, not redundant with it.
    \item \textbf{Vanilla LLM}, despite lacking any memory component, consistently improves under skill augmentation across all four datasets. This is a particularly strong result: it confirms that the skill file surfaces user-specific decision heuristics that a universal system prompt fundamentally cannot capture, regardless of the quality of retrieved context.
\end{itemize}

\textsc{Sager}'s per-user skill evolution is a \textbf{model-agnostic plug-in}: it delivers consistent, additive gains regardless of whether the host agent uses static profiles, dynamic profiles, or no memory at all. The improvement magnitude correlates with interaction density, reinforcing the StatInit density hypothesis. These results establish that policy evolution is not a MemRec-specific optimization but a general-purpose capability applicable to any LLM-based recommendation agent.

\subsection{Ablation Study (RQ3)}
\label{sec:ablation}

\paragraph{Component Ablation.}
\textsc{Sager} integrates four interdependent designs: the policy skill itself, statistical initialization, listwise reasoning, and incremental CoT evolution. A natural question is whether the observed gains stem from their synergy or from one dominant component. To answer this, we perform a leave-one-out ablation: each row in Table~\ref{tab:ablation_skill} removes exactly one component while keeping the remaining three intact, isolating its marginal contribution under otherwise identical conditions.

\begin{table*}[h!]
\centering
\caption{Policy skill system ablation (four datasets, H@1/H@3/H@5).}
\label{tab:ablation_skill}
\resizebox{\textwidth}{!}{%
\begin{tabular}{l|ccc|ccc|ccc|ccc}
\toprule
\multirow{2}{*}{\textbf{Configuration}} & \multicolumn{3}{c|}{\textbf{GoodReads}} & \multicolumn{3}{c|}{\textbf{Yelp}} & \multicolumn{3}{c|}{\textbf{MovieTV}} & \multicolumn{3}{c}{\textbf{Books}} \\
 & \textbf{H@1} & \textbf{H@3} & \textbf{H@5} & \textbf{H@1} & \textbf{H@3} & \textbf{H@5} & \textbf{H@1} & \textbf{H@3} & \textbf{H@5} & \textbf{H@1} & \textbf{H@3} & \textbf{H@5} \\
\midrule
\rowcolor{gray!10}
\textbf{\textsc{Sager} (Full)} & \textbf{0.459} & \textbf{0.700} & \textbf{0.827} & \textbf{0.523} & \textbf{0.722} & \textbf{0.814} & \textbf{0.628} & \textbf{0.832} & \textbf{0.909} & \textbf{0.521} & \textbf{0.705} & \textbf{0.810} \\
\midrule
w/o Policy Skill (full removal) & 0.399 & 0.666 & 0.805 & 0.486 & 0.691 & 0.791 & 0.588 & 0.782 & 0.882 & 0.511 & 0.692 & 0.801 \\
w/o Statistical Init           & 0.441 & 0.685 & 0.817 & 0.506 & 0.710 & 0.804 & 0.609 & 0.818 & 0.899 & 0.517 & 0.699 & 0.806 \\
w/o Listwise Reasoning          & 0.427 & 0.673 & 0.812 & 0.494 & 0.703 & 0.798 & 0.600 & 0.811 & 0.895 & 0.512 & 0.696 & 0.804 \\
w/o Incremental CoT             & 0.430 & 0.678 & 0.815 & 0.502 & 0.707 & 0.800 & 0.608 & 0.819 & 0.901 & 0.514 & 0.697 & 0.805 \\
\bottomrule
\end{tabular}%
}
\end{table*}

The component contribution ranking is \textit{consistent across all four datasets}: \textbf{w/o Skill $>$ w/o Listwise $>$ w/o CoT $\approx$ w/o StatInit}, revealing a clear internal hierarchy within the \textsc{Sager} architecture.

\begin{itemize}[leftmargin=*]

\item \textbf{Policy skill is the single most impactful component.}
On the three denser benchmarks, complete removal causes the largest drop: GoodReads $-13.1\%$, Yelp $-7.1\%$, MovieTV $-6.4\%$ on H@1. Consistent with the StatInit density hypothesis, GoodReads (highest density) shows the most pronounced degradation and Books the smallest ($-1.9\%$). The drop is concentrated on H@1, with H@3/H@5 degrading systematically less, reinforcing the asymmetric gain pattern: policy skills primarily improve \textit{top-rank decisional sharpness}.

\item \textbf{Listwise reasoning is critical across all benchmarks.}
Removing listwise (reverting to pointwise scoring) causes GoodReads $-7.0\%$, Yelp $-5.5\%$, MovieTV $-4.5\%$, Books $-1.7\%$ H@1 drops. This validates the Score-Tie Bottleneck analysis (\S\ref{sec:listwise}): without listwise's strict total ordering, the slim skill's tie-breaking function is neutralized by collapsed pointwise scores.

\item \textbf{StatInit and Incremental CoT form a complementary upstream-downstream pipeline.}
Both contribute meaningfully ($\sim$$3$--$4\%$ on most datasets, with CoT reaching $6.3\%$ on the densest benchmark GoodReads where iterative refinement benefits most): \textsc{StatInit} provides a frequency-grounded starting point; the CoT engine refines it into actionable decision strategy via differential updates. Neither alone is sufficient, but together they compound.

\end{itemize}

\paragraph{Incremental vs.\ Full-Replacement CoT.}
\label{sec:app_cot_ablation}

The component ablation above establishes that the CoT engine is essential, but does not reveal \textit{how} it should update the skill. Two natural strategies exist: (i)~\textit{full-replacement}, where the LLM rewrites the entire skill from the latest interaction, and (ii)~\textit{incremental}, where the LLM outputs only a structured diff $\Delta_u$ that is merged into the existing skill. To disentangle their effects, Table~\ref{tab:ablation_cot} compares four progressively stronger configurations: no policy skill (MemRec baseline), StatInit only (no CoT refinement), full-replacement CoT, and incremental CoT (ours).

\begin{table*}[h!]
\centering
\caption{Incremental vs.\ full-replacement CoT comparison (four datasets, H@1/H@3/H@5).}
\label{tab:ablation_cot}
\resizebox{\textwidth}{!}{%
\begin{tabular}{l|ccc|ccc|ccc|ccc}
\toprule
\multirow{2}{*}{\textbf{Configuration}} & \multicolumn{3}{c|}{\textbf{GoodReads}} & \multicolumn{3}{c|}{\textbf{Yelp}} & \multicolumn{3}{c|}{\textbf{MovieTV}} & \multicolumn{3}{c}{\textbf{Books}} \\
 & \textbf{H@1} & \textbf{H@3} & \textbf{H@5} & \textbf{H@1} & \textbf{H@3} & \textbf{H@5} & \textbf{H@1} & \textbf{H@3} & \textbf{H@5} & \textbf{H@1} & \textbf{H@3} & \textbf{H@5} \\
\midrule
\rowcolor{gray!10}
\textbf{Incremental CoT (ours)} & \textbf{0.459} & \textbf{0.700} & \textbf{0.827} & \textbf{0.523} & \textbf{0.722} & \textbf{0.814} & \textbf{0.628} & \textbf{0.832} & \textbf{0.909} & \textbf{0.521} & \textbf{0.705} & \textbf{0.810} \\
Full-Replacement CoT   & 0.435 & 0.689 & 0.820 & 0.507 & 0.713 & 0.807 & 0.613 & 0.824 & 0.904 & 0.518 & 0.700 & 0.807 \\
StatInit Only (no CoT) & 0.430 & 0.678 & 0.815 & 0.502 & 0.707 & 0.800 & 0.608 & 0.819 & 0.901 & 0.514 & 0.697 & 0.805 \\
No Policy Skill (MemRec) & 0.400 & 0.666 & 0.805 & 0.487 & 0.691 & 0.791 & 0.588 & 0.782 & 0.882 & 0.511 & 0.692 & 0.801 \\
\bottomrule
\end{tabular}%
}
\end{table*}

Incremental CoT outperforms full-replacement by an average $+2.8\%$ on H@1, while full-replacement barely improves over StatInit-only ($+1.0\%$ on average). The diagnosis is clear: full-replacement CoT rewrites the entire skill from a single interaction, discarding the high-confidence frequency prior accumulated by \textsc{StatInit} and reducing the policy to a single-round snapshot. Incremental CoT, by outputting only the structured diff ($\Delta_u$), preserves statistical stability and stacks each interaction's signal precisely atop the existing prior. This result provides direct empirical support for the \textit{refine-not-replace} principle introduced in \S\ref{sec:cot_evolution}.

\subsection{Design Sensitivity Analysis (RQ4)}
\label{sec:skill_analysis}

In this section, we examine \textit{how sensitive} \textsc{Sager} is to two key hyperparameters that govern the policy skill lifecycle: the number of warmup rounds $\tau$ (controlling evolution depth) and the slim skill injection length (controlling information bandwidth at inference).

\paragraph{Effect of Warmup Rounds $\tau$.}
The warmup phase transforms the statistical prior from \textsc{StatInit} into a fully personalized policy skill via repeated contrastive CoT refinement. Too few rounds may leave the skill under-developed; too many risk over-fitting to limited interaction data. To locate the optimal operating point, Figure~\ref{fig:warmup_tau} sweeps $\tau \in \{0,1,2,3\}$ on Yelp, MovieTV, and Books (1k users per dataset), reporting H@1, H@3, and H@5. Three findings emerge from this sweep:

\begin{figure*}[t]
    \centering
    \includegraphics[width=\textwidth]{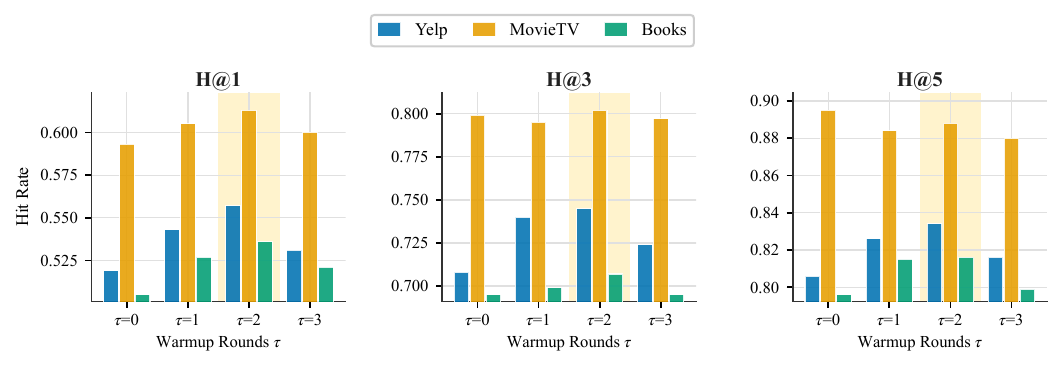}
    \vspace{-3mm}
    \caption{Effect of warmup rounds $\tau$ on H@1, H@3, and H@5 (one panel per metric; bar colors distinguish datasets). $\tau=2$ achieves the best performance across most metrics and datasets (yellow highlight).}
    \label{fig:warmup_tau}
\end{figure*}

\begin{itemize}[leftmargin=*]

\item \textbf{$\tau=1$ already yields substantial H@1 gains} (Yelp $+4.6\%$, MovieTV $+2.0\%$, Books $+4.4\%$ over $\tau{=}0$), but H@3/H@5 gains remain limited. This indicates that a single round of contrastive reasoning primarily sharpens top-rank precision, consistent with the asymmetric gain pattern.

\item \textbf{$\tau=2$ achieves the best results across most metrics simultaneously.}
The explanation is mechanistic: two warmup rounds complete the full \textsc{Discover}$\to$\textsc{Reinforce} cycle. The first round discovers new preference signals and appends them at confidence \texttt{low}; the second round confirms them and promotes to \texttt{medium} (injection-eligible). A single round can only discover or reinforce, but not both, for any given preference.

\item \textbf{$\tau=3$ regresses}, most pronounced on sparse-history Books (H@3 drop to $\tau{=}0$ levels). The third-round interaction heavily overlaps with already-confirmed preferences, triggering excessive \textsc{Weaken} operations on fragile signals. This suggests that over-evolution with limited data risks destabilizing the prior rather than refining it.

\end{itemize}

\noindent Based on these results, \textbf{$\tau=2$ is adopted as the default} across all datasets and metrics.

\paragraph{Injection Length and Cognitive Injection Boundary.}
The injection length controls how much of the evolved skill the LLM actually \textit{sees} at inference time. The Cognitive Injection Boundary (\S\ref{sec:skill_injection}) predicts a non-monotonic relationship: too little context under-specifies the user's policy, while too much triggers attention dilution. To test this prediction, Figure~\ref{fig:injection_length} sweeps five injection lengths (10, 30, 80, 150, 200 tokens) on 1k users per dataset, with the no-injection baseline MemRec as reference. Three distinct operating regimes emerge:

\begin{figure*}[t]
    \centering
    \includegraphics[width=\textwidth]{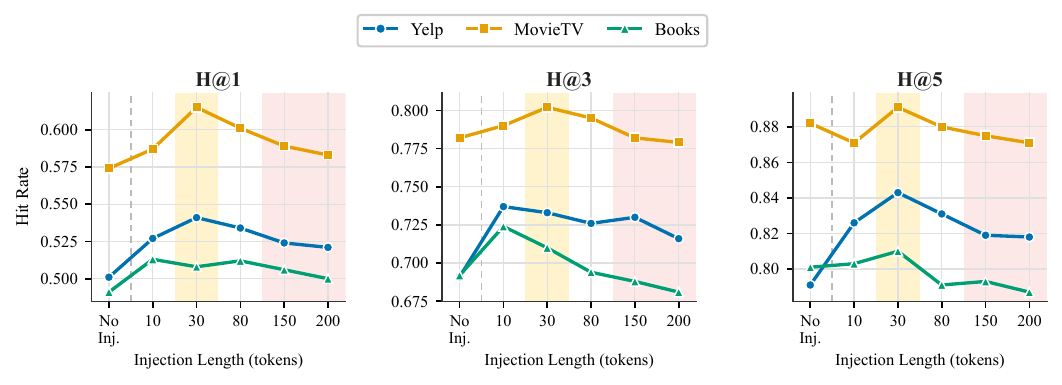}
    \vspace{-3mm}
    \caption{Effect of slim skill injection length on H@1, H@3, and H@5 (one panel per metric; line colors distinguish datasets). Yellow region: Cognitive Injection Boundary (30 tokens); red region: over-injection zone ($\geq$150 tokens).}
    \label{fig:injection_length}
\end{figure*}

\begin{itemize}[leftmargin=*]

\item \textbf{Under-injection (10 tokens):} H@1 already exceeds the no-injection baseline on all datasets, confirming that even minimal policy context provides discriminative value. However, the improvement is sub-optimal because ultra-short skills cannot encode both \texttt{likes} and \texttt{style} fields with sufficient specificity.

\item \textbf{Optimal zone (30 tokens):} This is the Pareto-optimal point. H@1 peaks on Yelp and MovieTV, and achieves robust performance across all datasets without systematic degradation on any metric. The 30-token budget aligns precisely with the slim skill's structured format (2--3 \texttt{likes} themes + 1 \texttt{style} phrase), suggesting that the Cognitive Injection Boundary reflects a structural constraint rather than an arbitrary threshold.

\item \textbf{Over-injection ($\geq$150 tokens):} H@1 declines monotonically while H@5 degrades more gradually, consistent with the \textit{attention dilution} hypothesis: excessively long injections introduce population-generic terms that crowd out the immediate instruction and collaborative signals, with top-rank precision suffering first. This confirms the Injection Paradox: richer does not mean better.

\end{itemize}

Based on these results, \textbf{30 tokens} is adopted as the default injection length.

\section{Related Work}
\label{sec:related_work}

\paragraph{Memory Architectures for LLM Agents.}
Building autonomous agents capable of long-horizon tasks requires overcoming LLM context window constraints~\cite{liu2024lost} and ensuring long-term knowledge retention. Early solutions combined LLMs with external vector databases~\cite{johnson2019billion} to create RAG pipelines~\cite{lewis2020retrieval}; Graph RAG~\cite{edge2024local} further demonstrates the value of structuring retrieved context into knowledge graphs. Dedicated memory systems then emerged: MemGPT~\cite{packer2023memgpt} introduces OS-inspired virtual context management, Zep~\cite{rasmussen2025zep} structures memory into temporal knowledge graphs, and Generative Agents~\cite{park2023generative} demonstrate how reflective synthesis drives believable agent behavior. Learning-based memory policies~\cite{xu2025amem, yan2025memory} further train dedicated managers to optimize storage and retrieval. However, these systems target conversational or factual domains, neglecting the specialized high-order connectivity required for collaborative recommendation.

\paragraph{Memory in Agentic Recommendation.}
Memory integration in recommendation has evolved from latent states in sequential models~\cite{hidasi2015session, kang2018self} to explicit dynamic structures managed by LLM agents. Early work explored stateless prompting~\cite{liu2023chatgpt, lyu2024llm, geng2022recommendation, bao2023tallrec}. Subsequently, iAgent~\cite{xu2025iagent} introduced static user profiles and Chat-REC~\cite{gao2023chat} adopted RAG-style history retrieval, but neither can adapt to real-time feedback. To address plasticity, recent work introduces dynamic memory with planning or tool-use~\cite{wang2024recmind, huang2025recommender, wang2024macrec, shu2023rah}: i$^2$Agent~\cite{xu2025iagent} and RecBot~\cite{tang2025interactive} employ self-reflection after each interaction; simulation frameworks AgentCF~\cite{zhang2024agentcf}, Agent4Rec~\cite{zhang2024generative}, and RecAgent~\cite{wang2025user} model users and items as agents with evolving memories. MemRec~\cite{chen2026memrec} advances beyond isolated self-reflection by introducing a collaborative memory graph that propagates semantic signals across users. PersonaX~\cite{shi2025personax} explores user modeling for long behavior sequences. More recently, STAR~\cite{wu2026internalizing} distills multi-agent teacher reasoning into a single model via SFT+GRPO, and RecNet~\cite{li2026recnet} implements self-evolving preference propagation inspired by network routing. \textit{All} these systems evolve \textit{what the agent knows} (memory), but \textit{none} evolves \textit{how the agent reasons} (policy), a fundamental gap that \textsc{Sager} addresses.

\paragraph{Self-Evolving and Self-Improving Agents.}
A nascent line of work pursues autonomous self-improvement at the system scale~\cite{cai2025agentic, wang2026self}. Voyager~\cite{wang2023voyager} maintains a skill library for Minecraft exploration; SkillRL~\cite{xia2026skillrl} co-evolves a skill bank and agent policy via recursive reinforcement learning, while the Agent Skills survey~\cite{xu2026agent} provides a taxonomy of composable skill architectures. AgenticRS~\cite{hu2026rethinking} proposes a blueprint for transforming static recommendation pipelines into self-evolving agentic systems. In recommendation, \citet{wang2026self} demonstrate system-level autonomous model optimization deployed at YouTube scale. \textsc{Sager} operates at a complementary granularity: per-user policy evolution, accumulating individualized decision knowledge that system-level evolution cannot provide. Rather than evolving the \textit{model} or \textit{data}, \textsc{Sager} evolves the \textit{reasoning strategy} for each individual user.

\paragraph{Knowledge Injection and Chain-of-Thought Reasoning.}
Injecting external knowledge into LLMs has been studied via retrieval augmentation~\cite{lewis2020retrieval}, tool use~\cite{zhao2024let, Chase_LangChain_2022}, and structured prompting. Chain-of-thought prompting~\cite{wei2022chain} improves general reasoning; self-consistency~\cite{wang2022self} and tree-of-thought~\cite{yao2023tree} further explore structured inference, while listwise reranking~\cite{hou2024large} demonstrates the superiority of comparative over absolute scoring, a property that \textsc{Sager}'s skill-augmented listwise reasoning directly exploits. ReRec~\cite{huang2026rerec} augments LLM recommendation reasoning via reinforcement fine-tuning with graph-based dual rewards. Yet none of these methods personalizes the reasoning \textit{strategy} per user. \textsc{Sager} is, to our knowledge, the first to treat the recommendation policy as a per-user, self-evolving natural-language artifact, bridging the gap between static knowledge injection and adaptive reasoning personalization.

\section{Conclusion}
\label{sec:conclusion}
This paper identifies a structural limitation shared by all existing LLM-based recommendation agents: while these systems personalize \textit{what} the agent knows through evolving per-user memory, \textit{how} the agent reasons remains a universal, static policy shared identically across all users. We formally distinguish policy evolution from memory evolution and model evolution, establishing it as a previously unaddressed dimension of personalized recommendation.
To close this gap, we propose \textbf{\textsc{Sager}}, the first recommendation agent framework in which each user is equipped with a dedicated, self-evolving \textit{policy skill}. \textsc{Sager} operates through a four-stage closed loop: retrieving collaborative memory from a semantic graph, extracting a minimal-footprint slim skill that respects the empirically discovered \textit{Cognitive Injection Boundary}, performing skill-augmented listwise ranking that eliminates the \textit{Score-Tie Bottleneck} of pointwise approaches, and evolving the policy skill via an incremental contrastive chain-of-thought engine that diagnoses reasoning flaws through the contrast between accepted and unchosen items. A dual-level initialization architecture ensures graceful cold-start, while the \textit{refine-not-replace} evolution principle preserves statistical stability across update cycles.
Experiments across four benchmarks confirm state-of-the-art performance, with particularly strong H@1 gains concentrated at the decision boundary where candidate scores are near-tied. Ablation studies demonstrate that each component contributes independently, and that policy evolution produces gains orthogonal to memory accumulation, confirming that personalizing \textit{how} the agent reasons, not merely \textit{what} it knows, constitutes a qualitatively distinct source of recommendation improvement.

\bibliographystyle{plainnat}
\bibliography{custom}

\clearpage
\appendix
\appendix
\section{Experimental Setup and Implementation Details}
\label{sec:appendix_implementation}

\subsection{Dataset Details}
\label{app:dataset_details}

We utilize four datasets widely used in recommendation research, encompassing diverse domains such as e-commerce, social reading, entertainment, and local services. As mentioned in Section \ref{sec:experiments}, we adopt the versions of these datasets augmented with natural language user instructions from \textbf{InstructRec} \cite{xu2025iagent}. The original data sources and their detailed descriptions are provided below:

\paragraph{Books.}
Derived from the \textbf{Amazon review dataset}\footnote{\url{https://cseweb.ucsd.edu/~jmcauley/datasets/amazon_v2/}} \cite{ni2019justifying}, this subset focuses on book recommendations. It is characterized by incredibly sparse interactions and a vast item space. User preferences in this domain are typically stable and highly content-driven, focusing on specific genres, authors, or themes.

\paragraph{Goodreads.}
Collected from the \textbf{Goodreads} social book cataloging website\footnote{\url{https://cseweb.ucsd.edu/~jmcauley/datasets/goodreads.html}} \cite{wan2019fine}, this dataset is notably dense compared to others. It features strong community interactions and rich metadata about books, including series information. Users on Goodreads often exhibit series-aware reading behaviors and are influenced by social signals.

\paragraph{MovieTV.}
Also originating from the \textbf{Amazon review dataset} \cite{ni2019justifying}, this dataset covers movies and TV shows. The domain is marked by volatile user preferences often influenced by immediate context or trending content. While metadata like genre and cast are important, item recency frequently plays a critical role in user decision-making.

\paragraph{Yelp.}
Sourced from the \textbf{Yelp Dataset}\footnote{\url{https://www.kaggle.com/datasets/yelp-dataset/yelp-dataset}}, this dataset consists of reviews for local businesses like restaurants and services. It is characterized by strong categorical constraints (e.g., cuisine type) and the critical importance of attributes like price range and location. User preferences here are often highly context-dependent.

\subsection{Baseline Model Details}
\label{app:baseline_details}

This appendix provides detailed descriptions of the baseline models used in our comparative evaluation. Following the categorization in the main text, we group these baselines based on their underlying memory paradigms into two major categories: traditional pre-LLM methods using latent embeddings, and memory-based approaches developed in the post-AgentRS era using semantic memory.

\subsubsection{Traditional Pre-LLM Methods (Latent Embeddings)}

These models represent the conventional paradigm where historical information is encoded and preserved using dense latent vectors, without explicit semantic memory structures for reasoning agents.

\begin{itemize}[leftmargin=*]
    \item \textbf{LightGCN} \cite{he2020lightgcn}: A state-of-the-art graph collaborative filtering model that simplifies the Graph Convolutional Network (GCN) design by removing feature transformation and nonlinear activation. It learns user and item embeddings by linearly propagating them on the user-item interaction graph, capturing high-order collaborative signals through structural connections.

    \item \textbf{SASRec} \cite{kang2018self}: A leading sequential recommendation model based on the self-attention mechanism. It models the entire user sequence to capture long-term semantics and dynamic dependencies, using an attention mechanism to selectively focus on relevant items in the history for making predictions.

    \item \textbf{P5} \cite{geng2022recommendation}: A unified framework that formulates various recommendation tasks as sequence-to-sequence language modeling problems. It utilizes a pre-trained T5 backbone and represents users and items as sequence tokens (IDs) within personalized prompts. While LLM-based, the original P5 relies on pre-trained knowledge related to these IDs and does not incorporate an evolving, descriptive memory component.
\end{itemize}

\subsubsection{Memory-based Approaches (Post-AgentRS Era)}

This category encompasses approaches developed in the era following AgentRS, utilizing LLMs with varying degrees of semantic memory capabilities.

\paragraph{(1) Models with No Explicit Memory.}
These models operate by directly processing raw interaction histories without maintaining a persistent, structured semantic memory store.

\begin{itemize}[leftmargin=*]
    \item \textbf{Vanilla LLM (Zero-Shot Prompting)} \cite{liu2023chatgpt}: This baseline represents the direct application of a powerful instruction-tuned LLM (e.g., GPT-4o-mini) via API calls. For each prediction, the user's entire sequence of historical interactions is converted into a natural language string and fed into the LLM as a static context prompt. The model performs zero-shot selection from candidate items based solely on this provided raw history, serving as a baseline to measure the LLM's inherent capabilities independent of designed memory architectures.
\end{itemize}

\paragraph{(2) Static Memory Agents.}
These agents utilize descriptive semantic information about users and items, but this "memory" remains fixed as a static context during inference and does not evolve.

\begin{itemize}[leftmargin=*]
    \item \textbf{iAgent} \cite{xu2025iagent}: An LLM-based autonomous agent designed for recommendation. It employs a static profile for each user, constructed from their historical interactions and available descriptive data. This fixed profile is fed into the LLM as context to generate recommendations. The key characteristic is that its understanding of the user does not adapt over time after initial construction.
\end{itemize}

\paragraph{(3) Dynamic Memory Agents (Isolated Updates).}
These agents possess a dynamic memory mechanism, allowing them to reflect on interactions and update their understanding. However, these updates are isolated to the individual agent and do not propagate collaboratively.

\begin{itemize}[leftmargin=*]
    \item \textbf{i$^2$Agent} \cite{xu2025iagent}: An extension of iAgent that introduces a "reflection" mechanism. After recommendations, the agent can reflect on user feedback to refine its internal state or strategy for future interactions. While dynamic, these reflections are confined to the individual agent's experience with a specific user.

    \item \textbf{AgentCF} \cite{zhang2024agentcf}: An agent-based collaborative filtering framework that simulates user-item interactions. Agents representing users and items can autonomously interact, learn from these interactions, and update their own preferences or characteristics. The memory update is dynamic but remains localized to the individual agents involved in the direct interaction.

    \item \textbf{RecBot} \cite{tang2025interactive}: A conversational recommender system that uses an LLM to engage with users. It maintains a dynamic dialogue history and can update its understanding of user preferences based on the ongoing conversation. This dynamic memory allows for multi-turn interactions but is limited to the context of the current user session.
\end{itemize}

\paragraph{(4) Dynamic Collaborative Memory.}

\begin{itemize}[leftmargin=*]
    \item \textbf{MemRec} \cite{chen2026memrec}: A memory-augmented LLM agent for personalized recommendation that serves as our primary baseline. MemRec retrieves user context via semantic similarity and collaborative signals from a user-item interaction graph, and synthesizes structured memory facets for re-ranking. It operates with a universal, static reasoning policy. \textsc{Sager} extends this backbone by introducing a per-user self-evolving policy skill that personalizes not only what the agent knows, but how it reasons.
\end{itemize}

\subsection{Implementation Details}
\label{sec:implementation_details}

\paragraph{Model Deployment.}
All experiments are conducted using \texttt{gpt-4o-mini} accessed via a unified OpenAI-compatible API endpoint, with temperature set to $0.0$ for reproducibility. We use the same API endpoint and model for all pipeline stages (retrieval, re-ranking, skill evolution).

\paragraph{Hyperparameters.}
The core memory and retrieval hyperparameters follow the original MemRec~\cite{chen2026memrec} configuration to ensure fair comparison. All retrieval, pruning, and re-ranking parameters are kept identical to MemRec defaults. For skill evolution, the warmup rounds $\tau$ are searched over $\{0, 1, 2, 3\}$. The slim skill injection length is searched over $\{10, 30, 80, 150, 200\}$ tokens; based on the averaged trend across multiple independent runs, 30 tokens achieves the best overall performance across datasets (individual runs may show a smaller gap between 30 tok and 80 tok, but 30 tokens is the only setting with no systematic degradation across any dataset or truncation point), and is therefore used as the default; confidence tiers have three levels (\texttt{low}/\texttt{medium}/\texttt{high}), with only \texttt{medium} and above eligible for injection.




\section{Qualitative Case Study: Skill Evolution in Action}
\label{sec:appendix_case_study}

\subsection{Case A: Yelp --- Emerging Preference Discovery (User 0)}
\label{sec:app_case_a}

User~0 has extensive interactions in Las Vegas, primarily in the Restaurants category. \textbf{StatInit} ($\tau=0$, zero LLM cost) constructs the initial skill from interaction frequency statistics. Key preferences:

\begin{verbatim}
- Restaurants (Confidence: high, Source: confirmed)
- $$ price range (Confidence: high, Source: confirmed)
- casual ambience (Confidence: high, Source: confirmed)
- Las Vegas area (Confidence: high, Source: confirmed)
- premium tobacco products (Confidence: low, Source: emerging)
\end{verbatim}
Injected slim skill: \texttt{likes: Restaurants, Food, \$\$ price range, casual ambience}

\noindent\textbf{After $\tau=1$ warmup:} \texttt{high-rated venues (4+ stars)} confidence advances \texttt{low $\to$ medium} via the Reinforce phase, reflecting a positive interaction at a highly rated venue.

\noindent\textbf{After $\tau=2$ warmup:} \texttt{high-rated venues (4+ stars)} further advances to \texttt{high}. The Discover phase surfaces two latent preferences invisible in frequency statistics:
\begin{verbatim}
- Indian cuisine (Confidence: low, Source: emerging)
- tobacco shops (Confidence: low, Source: emerging)
\end{verbatim}
These dimensions emerge only through contrastive CoT analysis of interaction context.

\subsection{Case B: Books --- Domain-Adaptive StatInit (User 0)}
\label{sec:app_case_b}

Books dataset has only free-text \texttt{title} and \texttt{description} fields. The domain-adaptive parser extracts genre/mood signals:

\begin{verbatim}
- mystery (Confidence: high, Source: confirmed)
- romance (Confidence: low, Source: confirmed)
- fantasy (Confidence: low, Source: confirmed)
- uplifting reads (Confidence: low, Source: confirmed)
\end{verbatim}
\textbf{EVOLVABLE Strategy:} \texttt{Must Include: Books in core genres: mystery, romance, fantasy. Tie Breaker: New releases or highly-rated books in core genres.}

Injected slim skill: \texttt{likes: mystery, romance, fantasy}. During $\tau=1$ evaluation (instruction: \textit{``...feel-good and predictable narrative...''}), the ground-truth item ranks \textbf{1st} (Hit@1), confirming the slim skill's genre signal effectively guided the listwise ranking decision.

\section{Prompt Templates}
\label{sec:app_prompts}

\subsection{Incremental CoT Evolution Prompt}
\label{sec:app_cot_prompt}

The incremental CoT evolution follows a three-phase \textit{refine-not-replace} protocol. Rather than regenerating the full skill from a single interaction (which destroys the statistical prior), it outputs only a differential update: \texttt{new\_preferences}, \texttt{reinforced}, and \texttt{weakened}. Below is the core template (abbreviated from \texttt{user\_strategy\_cot\_incremental.md}):

\begin{promptbox}[Incremental CoT Update Template (Abbreviated)]
\textbf{Purpose:} Incrementally update an existing user skill based on a new interaction. Output ONLY the changes (additions, reinforcements, and weakening of existing preferences).

\textbf{Key Principle:} The existing skill was built from long-term history statistics. A single new interaction should REFINE it, not REPLACE it. Preserve high-confidence long-term preferences.

\textbf{Phase 1 (Reinforce):} Does this interaction REINFORCE any existing preferences?
- Which existing Core Preferences are supported by this choice?

\textbf{Phase 2 (Discover):} Does this interaction REVEAL new preferences not in the existing skill?
- What new signals emerge? Are they exploratory or consistent with existing patterns?

\textbf{Phase 3 (Weaken):} Does this interaction WEAKEN any existing preferences?
- Only if the positive choice clearly contradicts an existing preference.
- Be conservative: a single interaction rarely invalidates a long-term pattern.

\textbf{Output:}
\{"analysis": "2-3 sentences reasoning",
 "incremental\_update": \{
   "new\_preferences": [\{"attribute": ..., "reason": ...\}],
   "reinforced": [\{"attribute": ..., "evidence": ...\}],
   "weakened": [\{"attribute": ..., "reason": ...\}]
 \}\}

\textbf{Constraints:} (1) Conservative---prefer reinforcing over weakening; (2) Specific---concrete attributes only; (3) No Must-Avoid rules from a single non-selection.
\end{promptbox}

\subsection{Slim Profile Extraction Prompt}
\label{sec:app_slim_extract}

The slim profile extractor compresses the full skill repository to a $\sim$30-token injection representation, using only signals with confidence \texttt{medium} or above:

\begin{promptbox}[Slim Profile Extraction Prompt (Core)]
\textbf{Purpose:} Convert verbose user memory into a minimal ($\sim$30 tokens) structured profile for tie-breaking in ranking.

Extract:
1. \textbf{likes}: Top 2--3 preference themes most strongly distinguishing this user.
2. \textbf{style}: One phrase describing their overall taste pattern.

Only include patterns with Confidence: medium or high. Output must be $\leq$50 tokens total.

\textbf{Output format:} \texttt{likes: X, Y, Z | style: [decision style phrase]}
\end{promptbox}

\subsection{Listwise Ranking Prompt}
\label{sec:app_listwise_prompt}

The listwise ranking prompt instructs the LLM to produce a complete ordering rather than independent scores. Below is the core structure:

\begin{promptbox}[Listwise Ranking Prompt (Core Structure)]
You are an expert personalized recommendation system.

\textbf{User's Current Request:}
\textcolor{promptkeyword}{\{instruction\}}

\textbf{User Preferences (Memory):}
\textcolor{promptkeyword}{\{formatted\_facets\}}

\textbf{Per-User Preference Skill:}
\textcolor{promptkeyword}{\{slim\_skill\}}
*(Use as tie-breaker only when candidates are equally relevant.)*

\textbf{Candidate Items:}
\textcolor{promptkeyword}{\{formatted\_candidates\}}

\textbf{Task:} Rank ALL candidate items from most to least relevant. Output a JSON array of item IDs in ranked order, with a brief rationale for each.

Output format: \{"ranking": [\{"item\_id": ..., "rationale": "..."\}, ...]\}
\end{promptbox}

This formulation eliminates the score-tie problem entirely: every item receives a unique rank position, and the LLM leverages its comparative reasoning strength rather than struggling with absolute calibration.

\subsection{Memory Retrieval Prompt}
\label{sec:app_memory_retrieval_prompt}

The memory retrieval and update components of \textsc{Sager} are implemented on top of an existing memory-based recommendation agent (MemRec~\cite{chen2026memrec} in our experiments). \textsc{Sager} is designed to be agent-agnostic: the policy skill can be plugged into any memory recommendation agent that exposes per-user memory and collaborative neighbor context. The memory retrieval prompt below is inherited from the underlying agent and is shown for completeness; it synthesizes per-user memory and collaborative neighbor signals into structured preference facets that are subsequently used in the listwise ranking stage.

\begin{promptbox}[Memory Retrieval Prompt (Core Structure)]
You are an intelligent memory retrieval system for personalized recommendation. Your task is to analyze the user's personal memory and collaborative memories from their neighbors to extract preference facets.

\textbf{Target User:} User \textcolor{promptkeyword}{\{user\_id\}}

\textbf{User's Personal Memory:}
\textcolor{promptkeyword}{\{user\_mem\_bullets\}}

\textbf{Collaborative Neighbor Memories:}
The following neighboring users and items provide collaborative signals for understanding this user's preferences:
\textcolor{promptkeyword}{\{neighbor\_table\_json\}}

\textbf{Task:} Identify \textcolor{promptkeyword}{\{n\_facets\}} distinct preference facets. For each facet, provide: (1) a concise natural language description; (2) a confidence score in $[0,1]$; (3) a list of supporting neighbor IDs.

\textbf{Output format:}
\{"facets": [\{"facet": ..., "confidence": ..., "supporting\_neighbors": [...]\}, ...],
 "support\_edges": [\{"from": ..., "to": ..., "w": ...\}, ...]\}
\end{promptbox}

\section{Global Skill Example: Yelp Dataset}
\label{sec:app_global_skill}

The global skill $\mathcal{S}_G$ provides a population-level prior for users with no interaction history (cold start, Level 1 of the dual-level initialization). It is synthesized once per dataset by a meta-LLM from domain statistics, and is inherited as the starting point before any per-user CoT evolution. Below is the full global skill for the Yelp dataset.

\begin{promptbox}[Global Skill: Yelp Dataset (full)]
\textbf{Persona Clusters (interaction-based):}
\begin{itemize}[leftmargin=*, itemsep=0pt, topsep=2pt]
  \item Light Reviewers ($<$15 interactions, $\sim$30\%): limited signal; rely on popularity/location.
  \item Moderate Reviewers (15--50 interactions, $\sim$50\%): emerging category preferences.
  \item Heavy Reviewers ($>$50 interactions, $\sim$20\%): stable taste profiles, neighborhood regulars.
\end{itemize}

\textbf{Interest Clusters:} Cuisine-Loyal | Price-Sensitive | Neighborhood-Bound | Experience-Seeker.

\textbf{Strategy 1 --- Guided Discovery} (Light Reviewers):
Prioritize high-rating local businesses; restrict to 2--3 dominant categories; apply geo-proximity filter; avoid niche or low-review establishments.

\textbf{Strategy 2 --- Preference Reinforcement} (Moderate Reviewers):
70\% from confirmed cuisine/category facets; 30\% from adjacent categories; maintain price-tier consistency.

\textbf{Strategy 3 --- Exploratory Recommendation} (Heavy Reviewers, high category spread):
50\% established facets; 50\% novel/adjacent; 5--7 categories; cross-district exploration enabled.

\textbf{Strategy 4 --- Comfort Zone} (Heavy Reviewers, low category spread):
85\% from top-3 confirmed category/price facets; 15\% closely related variants.

\textbf{Dynamic Adjustment Rules:}
\begin{itemize}[leftmargin=*, itemsep=0pt, topsep=2pt]
  \item Recency weighting: last 3 interactions $1.8\times$; interactions 4--15 $1.0\times$; $>$15 $0.6\times$.
  \item If $>$60\% of neighborhood-category businesses visited: expand geo-radius by one zone.
  \item Category skipped $\geq$3 consecutive recommendations: suppress for 10 cycles.
\end{itemize}

\textbf{Memory Integration:}
Prioritize item-based CF over user-based CF given the relatively sparse interaction graph. Neighbor agreement $>$65\%: boost facet confidence $+$20\%. Leverage Yelp category taxonomy to generalize sparse interactions.

\textbf{Default priority order:} accuracy $\succ$ locality $\succ$ diversity $\succ$ novelty.
\end{promptbox}
\section{Domain-Aware Statistical Initialization: Extended Analysis}
\label{sec:app_statinit_domain_aware}

\subsection{Cross-Domain Metadata Structure and Parser Design}
\label{sec:app_statinit_diagnosis}

The \textsc{StatInit} component must handle fundamentally different metadata structures across domains. Yelp provides richly structured metadata (categories, price tier, ambience tags, city, meal type, star ratings), while Books and MovieTV datasets contain only free-text \texttt{title} and \texttt{description} fields formatted as Python lists. Without domain-specific parsing, all regex patterns designed for Yelp return empty matches on Books/MovieTV, producing identical ``empty template'' skills for every user---completely defeating the purpose of statistical initialization.

We diagnose this failure empirically: on Books, the initial domain-agnostic implementation produced 100\% identical skill files (1.04KB Yelp-formatted templates), with \texttt{Core Preferences} sections populated with Yelp-specific terms (``Venues'', ``Ambience'', ``Location proximity'') irrelevant to book recommendation.

\subsection{Domain-Adaptive Parser Architecture}
\label{sec:app_statinit_parsers}

We introduce a \textbf{domain-adaptive multi-parser architecture} that selects the appropriate metadata extraction pathway based on dataset type:

\begin{table}[h!]
\centering
\caption{Domain-aware StatInit signal extraction. Each domain uses a specialized parser with dataset-appropriate dimensions.}
\label{tab:domain_parsers}
\resizebox{\columnwidth}{!}{%
\begin{tabular}{l|l|l|l}
\toprule
\textbf{Domain} & \textbf{Dimension 1} & \textbf{Dimension 2} & \textbf{Dimension 3} \\
\midrule
Yelp (baseline) & categories (regex) & price tier + city & ambience + meal type \\
Books/GoodReads & genres (15-class keyword) & favorite authors & reading moods (7-class) \\
MovieTV & genres (17-class + qualifier) & directors & viewing moods (8-class) \\
\bottomrule
\end{tabular}%
}
\end{table}

\paragraph{Key engineering decisions.}
\begin{itemize}[leftmargin=*, itemsep=1pt]
    \item \textbf{Description parsing}: \texttt{ast.literal\_eval} flattens Python-list-formatted descriptions before keyword extraction, recovering $\sim$300--800 characters of plot/review text per item.
    \item \textbf{Author/Director extraction}: restricted to \texttt{title + first paragraph ($\leq$400 chars)} to avoid false matches in plot descriptions (e.g., ``directed by fate'' vs.\ ``directed by Nolan'').
    \item \textbf{Noise filtering for MovieTV}: a \texttt{\_filter\_noise\_segments()} pass removes VHS/DVD physical condition descriptions (``VHS VG+'', ``Used: Good'', product dimensions) that pollute genre/mood extraction.
    \item \textbf{Signal pruning}: only signals consumed by downstream CoT or slim profile extraction are retained; series, reading level, and content format were dropped after chain-of-consumption analysis.
\end{itemize}

\subsection{Concrete Skill Examples: Books and MovieTV}
\label{sec:app_statinit_examples}

\paragraph{Books user skill (User~0, after StatInit):}
\begin{verbatim}
### Core Preferences
- mystery (Confidence: high, Source: confirmed)
- romance (Confidence: low, Source: confirmed)
- fantasy (Confidence: low, Source: confirmed)
- uplifting reads (Confidence: low, Source: confirmed)

### EVOLVABLE Strategy
Prioritization:
  Must Include: Books in core genres: mystery, romance, fantasy
  Tie Breaker: Award-winning or critically acclaimed in core genres
Exploration:
  Budget: 15%
  Directions: Adjacent genres with similar tone or themes
\end{verbatim}

\paragraph{Slim profile extracted from above:}
\begin{verbatim}
likes: mystery, romance, fantasy | style: complex world explorer
\end{verbatim}

\paragraph{MovieTV user skill (User~0, after StatInit):}
\begin{verbatim}
### Core Preferences
- action (Confidence: low, Source: confirmed)
- animation (Confidence: low, Source: confirmed)
- sci-fi (Confidence: low, Source: confirmed)
- funny tone (Confidence: medium, Source: confirmed)
- uplifting tone (Confidence: medium, Source: confirmed)

### EVOLVABLE Strategy
Ranking Criteria:
  Primary: Genre and director match to viewing history
  Secondary: Viewing mood (funny, uplifting)
  Tie Breaker: Highly-rated films in core genres
\end{verbatim}

\paragraph{Slim profile extracted from above:}
\begin{verbatim}
likes: action, animation, sci-fi, funny tone | style: genre variety seeker
\end{verbatim}



\end{document}